# The Economic Impact of Critical National Infrastructure Failure Due to Space Weather

Edward J. Oughton[1]

[1]Centre for Risk Studies, Judge Business School, University of Cambridge, Cambridge, UK

Corresponding author: Edward Oughton (e.oughton@jbs.cam.ac.uk)

# Summary and key words

Space weather is a collective term for different solar or space phenomena that can detrimentally affect technology. However, current understanding of space weather hazards is still relatively embryonic in comparison to terrestrial natural hazards such as hurricanes, earthquakes or tsunamis. Indeed, certain types of space weather such as large Coronal Mass Ejections (CMEs) are an archetypal example of a low probability, high severity hazard. Few major events, short time-series data and a lack of consensus regarding the potential impacts on critical infrastructure have hampered the economic impact assessment of space weather. Yet, space weather has the potential to disrupt a wide range of Critical National Infrastructure (CNI) systems including electricity transmission, satellite communications and positioning, aviation and rail transportation.

Recently there has been growing interest in these potential economic and societal impacts. Estimates range from millions of dollars of equipment damage from the Quebec 1989 event, to some analysts reporting billions of lost dollars in the wider economy from potential future disaster scenarios. Hence, this provides motivation for this article which tracks the origin and development of the socio-economic evaluation of space weather, from 1989 to 2017, and articulates future research directions for the field. Since 1989, many economic analyses of space weather hazards have often completely overlooked the physical impacts on infrastructure assets and the topology of different infrastructure networks. Moreover, too many studies have relied on qualitative assumptions about the vulnerability of Critical National Infrastructure. By modelling both the vulnerability of critical infrastructure and the socio-economic impacts of failure, the total potential impacts of space weather can be estimated, providing vital information for decision-makers in government and industry.

Efforts on this subject have historically been relatively piecemeal and this has led to little exploration of model sensitivities, particularly in relation to different assumption sets about infrastructure failure and restoration. Improvements could be expedited in this research area by open-sourcing model code, increasing the existing level of data sharing, and improving multi-disciplinary research collaborations between scientists, engineers and economists.

# Introduction

Space weather takes place when disturbances in Earth's upper-atmosphere and near-Earth space are capable of detrimentally affecting a wide range of key technologies (Hapgood et al. 2016), particularly Critical National Infrastructure (CNI) systems such as electricity transmission, satellite communications and positioning, aviation and rail transportation. Whereas the CNI impacts from terrestrial natural hazards (hurricanes, earthquakes etc.) have been widely studied, space weather has received little comparative attention. This is partially due to the low probability, high impact nature of this particular threat, and because attributing specific technological problems to space weather can be challenging. The 1859 Carrington event – named after Richard Carrington who observed the activity from his private observatory in South London – is the most popularised example of a space weather event. Although the Carrington event led to many issues on Earth, for example in the telegraph system, our technology has significantly progressed since the mid-19$^{th}$ century. Therefore, it is hard to determine how current CNI systems would respond to a modern-day Carrington event.

Growing recognition of this natural hazard is demonstrated by recent policy developments over the past decade aimed at bolstering national readiness to space weather threats. In the UK, there has been the release of a Space Weather Preparedness Strategy (Cabinet Office and BIS, 2015) and recognition of this threat on the National Risk Register of Civil Emergencies (Cabinet Office, 2017). On the other side of the Atlantic in the USA, President Obama signed on 13$^{th}$ October 2016 an Executive Order (13744) titled Coordinating Efforts to Prepare the Nation for Space Weather Events, which outlined the roles and responsibilities of different Federal government agencies in addressing the risks posed by space weather hazards.

While the study of space weather has progressed significantly, from scholarly recordings of astronomical events many centuries ago, to advanced modelling of how solar activity may drive geomagnetic disturbances on Earth, substantial efforts still need to be made in many areas of understanding. This ranges from fundamental scientific research, investigating the vulnerability of engineered systems, to assessing the potential socio-economic impacts due to CNI failure. While there has been ongoing research in the science and engineering domains for many decades, potential socio-economic impacts have received little attention. This has been identified as a shortcoming of space weather research for a decade (Lanzerotti, 2008) with little redress. Indeed, although there is newly invigorated desire to remedy this, there has been relatively little evidence published in the peer-reviewed literature. Additionally, there is little standardisation in the methods utilised leading to diverging results.

In light of these issues, this article first provides a general introduction to space weather before secondly considering the potential impacts on CNI. Thirdly, the evolution of the economic impact assessment literature on space weather is explored, before focusing on the strengths of the analysis undertaken to date and using this information to inform the future direction of the research area. Finally, conclusions are provided which reflect on how to enhance the current level of evidence on this matter.

## What is space weather?

Space weather can arise from many different types of eruptive phenomena associated with solar activity taking place on the surface of the sun (often referred to as a 'solar storm'). Consequently, it is the interaction of three primary forms of solar activity with Earth (or in near-Earth space) which causes space weather (Figure 2, later in this manuscript, provides an illustration of the potential impacts for each type of solar phenomena):

1. **Coronal Mass Ejections (CMEs)** are massive releases of billions of tonnes of charged particles and magnetic field from the surface of the sun (Webb and Howard, 2012).
2. **Solar Energetic Particle Events (SEPs)** are a huge increase in energetic particles, mainly of protons but also heavy ions, thrown out into space (Shea and Smart, 2012).
3. **Solar flares** are a rapid release of electromagnetic energy previously stored in inductive magnetic fields. Emitted radiation covers most of the electromagnetic spectrum, from radio waves to X-rays (Fletcher et al. 2011).

When occurring in combination the time line of impacts is likely to unfold as follows. Earth may first be bombarded with initial radiation (such as X-rays) from a solar flare approximately eight minutes after the event on the surface of the sun. A second barrage of very-high-energy solar particles (SEPs) may then arrive tens-of minutes later. Finally, a large CME may reach Earth somewhere between 1-4 days later depending on the speed of travel through the interplanetary space (see Liu et al. 2014). The magnetic field in the CME is likely to lead to a geomagnetic storm that could also last for multiple days as it drives huge electrical currents, especially at high geomagnetic latitudes, leading to bright auroral displays. Often two CMEs can be released in quick succession and analysis of past events suggest that this dual occurrence often leads to the most extreme impacts, as indicated by aurora occurring at low latitudes (Vaquero et al. 2008; Willis et al. 2005; Ribeiro, 2011).

Although extreme space weather events usually include all three of these solar phenomena, the most concern is associated with multiple very large and fast Carrington-sized Coronal Mass Ejections. CMEs are a key driver of coronal and interplanetary dynamics, particularly if a CME travelling across the interplanetary space then hits Earth in a southward magnetic field direction ($B_z$) as it can lead to the most dramatic interaction effect and therefore the largest geomagnetic disturbances (Webb and Howard, 2012). A southward directed CME interacts with the northward direction of Earth's magnetic field, leading to a cancelling effect, allowing CME energy to enter Earth's magnetic field. While auroras are often seen during modest forms of geomagnetic activity, they are generally enhanced by large CMEs, at which point the auroral band can expand equatorward to lower latitudes.

Aurora are caused by bands of charged particles being accelerated along Earth's magnetic field lines into the atmosphere, exciting atmospheric gases that then give off the light we see, as illustrated in Figure 1. Usually these visual displays occur in the auroral oval regions encircling Earth's poles and are indicative of geomagnetic activity. They can be altered in modest forms by the solar wind or in more extreme circumstances by a CME interacting with the planet's atmosphere.

Figure 1 Example of the aurora borealis in Yukon Territory, Canada ([source](#))

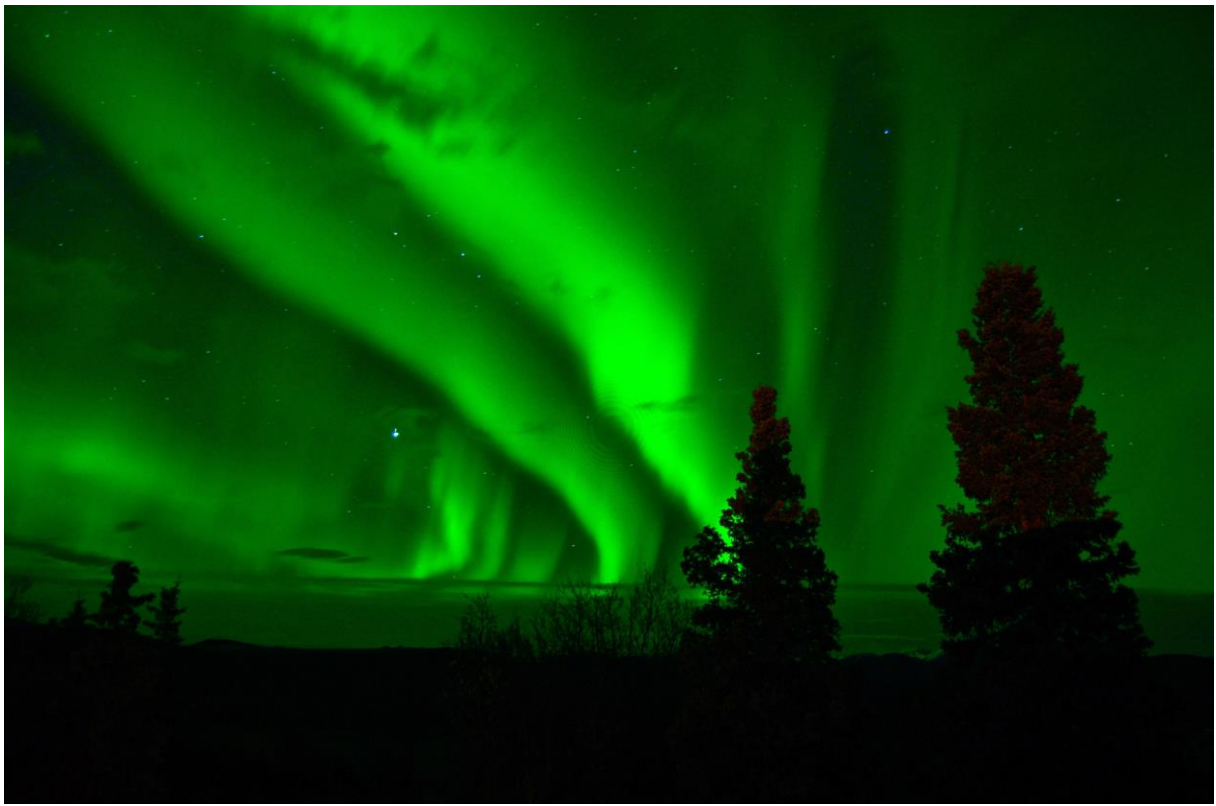

On average, the Sun's magnetic activity follows an 11-year solar cycle, with variable minimum and maximum periods. The latest solar cycle began in 2008 with minimal activity during the first few years. However, on July 23rd 2012 an extremely large CME estimated to be similar in size to the Carrington-event narrowly missed Earth (Baker et al. 2012). While the London 2012 Olympics was generally regarded as a success, had this CME hit Earth the day of the opening ceremony it could have been a very different story.

The strength and complexity of the Sun's magnetic field changes throughout the solar cycle, manifesting in visible 'sunspots' on the surface due to regions of concentrated magnetic field. The solar atmosphere changes during the sun cycle from a magnetically simple state to a complex configuration producing a larger number of 'sunspots', which are a key indicator of solar activity (Green and Baker, 2015). While there may be more activity during some parts of the solar cycle such as the declining phase (Juusola et al. 2015), solar eruptive phenomena are still the result of a random process. Therefore, there is potential for a significant space weather event to affect Earth at any time. Table 1 outlines a summarised list of major space weather events, some of the impacts caused and literature references for papers studying each case.

Table 1 A summary of the historical storm catalogue for major space weather events

| Year | Impact | Reference |
|---|---|---|
| 1847 | Spontaneous electrical currents observed in telegraph wires in the British Midlands, along railway corridors from Derby to Rugby, Birmingham, Leeds and Lincoln. | Barlow, 1848; Prescott, 1860; Cade, 2013 |
| 1859 | The archetypal example of space weather. Known as The Carrington Event, there was significant disruption caused to telegraph systems across the globe, and auroras were witnessed down to very low latitudes. | Boteler, 2006; Siscoe et al. 2006; Green and Boardsen, 2006; Ribeiro et al. 2011; Rodger et al. 2008; Saiz et al. 2016; Silverman, 2006; Tsurutani et al. 2003 |
| 1870 | A large storm produced aurora sightings in Lisbon and Coimbra (Portugal), Greenwich (UK), Munich (Germany) and Helsinki (Finland). | Vaquero et al. 2008 |
| 1872 | Aurora were sighted as low as 10-20° geomagnetic latitude, with significant recordings in Mumbai. | Moos, 1910a; 1910b; Uberoi, 2011 |
| 1882 | Strong aurora recorded in Scandinavia and North America. | Rubenson, 1882; Lewis, 1882 |

| 1921 | Similar in size to the Carrington Event, with significant GIC generated in Scandinavia. | Karsberg et al. 1959; Silverman and Cliver, 2001; Kappenman, 2006 |
|---|---|---|
| 1940 | Damage caused to the US telephone system and reported effects on the electricity network. | Harang, 1941; Davidson, 1940 |
| 1958 | Transatlantic communications were disrupted between Newfoundland and Scotland. There was a blackout in the Toronto area. | Anderson, 1978; Lanzerotti and Gregori, 1986 |
| 1989 | The Quebec power grid collapsed within 90 seconds. The well-documented Quebec power outage lasted nine hours. | Bolduc, 2002; Medford et al. 1989 |
| 2000 | The Bastille Day Event saw a very large CME and flare. | Tsurutani et al. 2005 |
| 2003 | The Halloween Storms included a mix of CMEs and flares leading to a one-hour power outage in Sweden. This storm also led to a radio blackout of high frequency communications, as well as disruption to GPS systems. | Pulkkinen et al. 2005; Tsurutani et al. 2005; Bergeot et al. 2010 |

Historical accounts record auroral sightings going back millennia, and ground-based magnetograph data has been recorded since the nineteenth century. However, it is only since the space age that we have been able to undertake large-scale digital monitoring of space weather events, approximately over the past 50 years. Although there is considerable focus on the most extreme space weather events such as a Carrington-level storm, it has been suggested that the potential economic impact of a prolonged period of moderate activity could be comparable to a single large incident (Schrijver, 2015). Indeed, research has suggested that just dealing with day-to-day space weather can pose a reliability challenge for electricity operators (for further detail see the work of Forbes and St. Cyr, 2008; 2012; 2017). Given the technological impacts associated with this storm catalogue, it is now pertinent to review the potential impacts to critical infrastructure.

## Impacts on CNI systems

Space weather has the capability to disrupt the critical technologies that comprise the national infrastructure system, although impacts vary by sector. This section provides an overview of the technologies potentially affected by space weather to help inform the economic impact assessment of this hazard. For a further summary of space weather worst-case environments see Hapgood et al. (2016), or for a detailed analysis of the impacts on engineered systems see Cannon et al. (2013). As illustrated in an impact tree in Figure 2, the aforementioned three key types of space weather can affect power grids, satellite systems, radio communications, aviation, rail transport, pipelines and undersea cables. Each of these impacts will be briefly discussed in this section.

Figure 2 Space weather impact tree (Adapted from Hapgood et al. 2016)

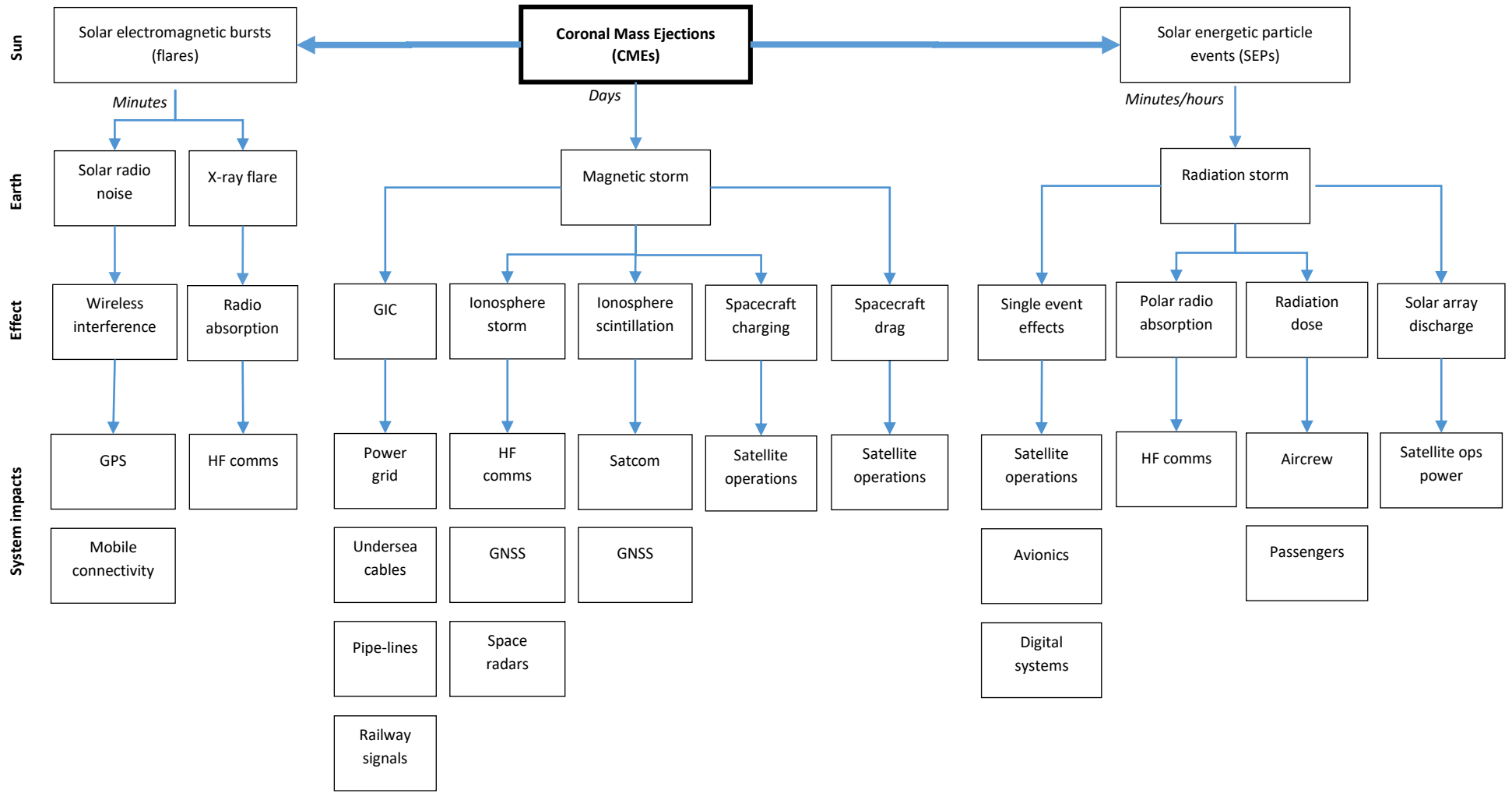

The largest focus has hitherto been on the threat to electricity transmission infrastructure, partly because energy underpins practically all daily activities and a prolonged loss of power would be disastrous for those affected. During a geomagnetic storm, Geomagnetically Induced Currents (GIC) are generated that are able to subject power grid assets to excessive thermal heating and voltage instability issues, potentially resulting in a loss of power. It is the speed of change in Earth's magnetic field that affects the generation of GIC, which can lead to immediate or cumulative damage in transformer components (Hutchins & Overbye, 2011). This rate-of-change of the magnetic field is best measured by d$B$/d$t$ which represents the time derivative of magnetic field variations on the ground (Kataoka and Ngwira, 2016). During a geomagnetic storm, there are many rapid global-scale variations in Earth's field and current systems which occur repeatedly. Known as 'substorms', these cause the most rapid changes in the magnetic field at the surface of Earth producing the largest GIC. Molinski (2002) states that this gives rise to half-cycle saturation in transformers, potential system voltage collapse, a loss in reactive power, as well as the generation of harmonics and excessive transformer heating. Geomagnetic latitude, ground conductivity and the power system network structure can influence the risk posed by this hazard. Figure 3 illustrates this including the damage pathway for electricity transmission infrastructure. Disruption to energy infrastructure can also lead to cascading failure, affecting other critical infrastructures such as transportation, digital communications and vital public health systems.

Figure 3 Detailed damage pathway for electricity transmission infrastructure (adapted from Boteler, 2015 and Samuelsson, 2013)

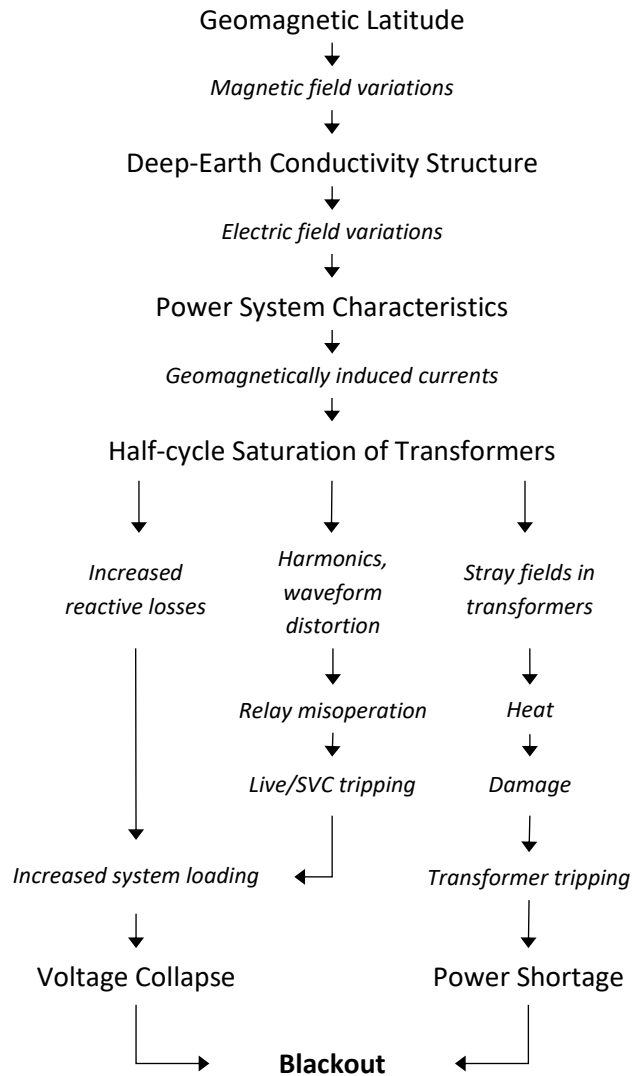

Considerable research activity has been taking place examining the impacts of geomagnetic activity on electricity transmission transformers, particularly in South Africa. This results from the geomagnetic activity spanning October and November 2003 (the 'Halloween Storms') which caused significant problems to many assets in the South African grid, leading to increased focus by electrical engineering researchers on these issues. See Gaunt and Coetzee (2007) and Gaunt (2014) for further detail, as well as Matandirotya et al. (2015) for an example of the GIC measurement and modelling research focusing on electricity transmission infrastructure in the country.

Spacecraft and satellites, including those enabling Global Positioning Systems (GPS), are susceptible to space weather phenomena particularly radiation bursts (Astafyeva et al. 2014). Indeed, a study of on-orbit spacecraft failures by Tafazoli (2009) found that over 10% of spacecraft anomalies were due to solar or magnetic storms. These impacts include both short-term and long-term effects. Firstly, during the event there may be problems with signal propagation and transmission due to interference caused by ionospheric storms and scintillation (Horne et al. 2013; Hapgood et al. 2016). Indeed, a loss of satellite capability can immediately disrupt many other infrastructure systems and economic sectors that rely on communications, navigation and timing services. On the other hand, long-term issues include spacecraft drag which can cause uncontrolled re-entry for satellites in low orbits or spacecraft charging issues which can affect onboard electronics. For satellites in higher orbits (e.g. geosynchronous), issues arise from both solar array damage and spacecraft charging (Hastings and Garrett, 2004, Garrett and Whittlesey, 2012; Lai and Tautz, 2006). Solar arrays can be degraded when cosmic rays and solar energetic particles penetrate them, as well as other electronic components (see Koons and Fennell, 2006). Power failures are often critical for spacecraft as 45% result in complete loss of spacecraft and 80% significantly affect the mission (Tafazoli, 2009).

In particular, high-frequency (HF) radio communications can be temporarily disrupted due to radio absorption (Neal et al. 2013; Rodger et al. 2008). During large electromagnetic radiation bursts released from the surface of the Sun, HF radio blackouts can occur lasting an hour or so, with the largest effects taking place in low latitude regions, where the Sun is highest in the sky. During SEPs, the greatest impacts can occur in the Polar Regions, sometimes lasting for several days. However, during periods of extreme space weather, aviation routes may need to be rerouted to avoid high latitude regions due to probable disruption to HF communications (Neal et al. 2013), as well as radiation risks to passengers and crew. Moreover, airline operations also suffer problems with avionics and GPS navigation systems during extreme events (Jones et al. 2005). This could potentially cause delays at major airports around the globe, particularly where there are a large number of flights that use the high latitude regions as an aviation corridor (e.g. New York to Tokyo, or Toronto to Hong Kong). New ambitions to send human missions to Mars and beyond also raise concerns about potential radiation exposure in space (Cougnet et al. 2004).

Rail transportation can be affected, but mainly from signals and tracks being subjected to GIC in two different ways. Firstly, signalling and train control system anomalies can occur during periods of high geomagnetic activity. Indeed, failures not related to recognised technical malfunctions can be up to seven times more likely on average (Ptitsyna et al. 2008; Eroshenko et al. 2010; Wik et al. 2009). Secondly, the structural integrity of rail infrastructure could also be affected if repeatedly subjected to extreme conditions, as this may increase the rate of corrosion. However, less evidence has been found to support the latter impact. This type of exposure is similar in nature to the risk to pipelines, particularly if there is cumulative long-term damage due to GIC as this can increase the likelihood of corrosion, shortening an asset's life (Pulkkinen et al. 2001a; Pulkkinen et al. 2001b; Gummow & Eng, 2002). Uncertainty exists regarding the magnitude of repeated exposure and the amount of time before an asset becomes affected, as exposure would not lead to immediate failure. Hence, it may not be possible to attribute the damage caused by exposure to high GIC to a space weather event, if the asset eventually fails many months or years after.

Effects to communications cables can take place in three different ways, including (i) the historical impacts on regional copper-wire electric telegraph and telephone systems, (ii) historical impacts on copper-wire transoceanic cables, and (iii) modern impacts on power systems in optical-fibre transoceanic cables. During the Carrington Event of 1859 many telegraph operators reported strange electrical effects. In fact, telegraph communications could still transmit and receive information even after systems were disconnected from the power supply due to GIC running through the cables. More recently, a documented case attempts to understand the problems caused by a large storm in 1958 whereby businesses and consumers in Finland were disrupted by the failures of two coaxial phone cable systems in the southern part of the country (Nevanlinna et al. 2001). The event was caused by blown fuses associated with the AC power supplies at repeater stations. Much like power grids, the submarine equivalent is equally affected by certain geographic and technical factors which, in this case, include the depth of the cable (Meloni et al. 1983). However, we have seen in recent decades a revolution in the technologies used to transmit information in digital communications networks. Modern systems therefore rely on fibre optics with glass fibre bring far less conductive than copper. Hence, far more at risk are the electrical cables which power fibre optic equipment (Medford et al. 1989). The following section will focus purely on the potential impacts to electricity transmission infrastructure as this has been the key subject of study for almost three decades.

## The evolution of the economic analysis of space weather

In this section, different chronological periods will be analysed based on how the study of the economic impacts of space weather has evolved over time. Prior to the geomagnetic storm of 1989 and the voltage collapse of the Hydro-Quebec electricity transmission grid, there were few major examples of critical infrastructure failure attributed to space weather events. Hence, there had been limited analysis of the consequential economic impacts. Therefore, we will track developments beginning after this event, with the analysis being broken down into three key temporal periods including 1989-2007, 2008-2013, and 2014-2017. The justification for these three chronological periods is provided in each subsection. The papers outlined in this part of the article have been summarised in Table 2.

Table 2 Summary of studies focusing on the economic impact assessment of space weather

| Year | Author | Infrastructure type | Geography | | Spatio-temporal impact | | Economic methodology | Economic impact | | | | Peer Reviewed? |
|---|---|---|---|---|---|---|---|---|---|---|---|---|
| | | | Country | Region | Population affected | Restoration period | | Asset damage | Direct economic impact | Indirect economic impact | Total economic impact | |
| 1990 | Barnes and Dyke | Electricity transmission infrastructure | USA | North East | Not stated | 50% connected in 16 hours, 75% in 24 hours, 100% in 48 hours | Value of Lost Load estimation | $16 million (1988 USD) | $3-6 billion (1988 USD) | Not modelled | Not modelled | Yes |
| 2002 | Bolduc | Electricity transmission infrastructure | Canada | Quebec | 9 million | N/A | Not stated | $13.2 million (Canadian dollars) | Not modelled | Not modelled | Not modelled | Yes |
| 2005 | Pulkkinen et al. | Electricity transmission infrastructure | Finland | Malmö | 50,000 | 1 hour | Not stated | Not stated | $0.5 million (USD) | Not modelled | Not modelled | Yes |
| 2008 | Kappenman (in Space Studies Board) | Electricity transmission infrastructure | USA | National assessment | Not stated | 4 to 10 years | Not stated | Not stated | $1-2 trillion (USD) | Not stated | Not stated | No |
| 2008 | Forbes and St. Cyr | Electricity transmission infrastructure | Multiple countries | National assessments | N/A | N/A | Econometrics | N/A | N/A | N/A | N/A | Yes |
| 2012 | Forbes and St. Cyr | Electricity transmission infrastructure | USA | National assessments | N/A | N/A | Econometrics | N/A | N/A | N/A | N/A | Yes |
| 2013 | Atmospheric Environmental Research for Lloyd's of London | Electricity transmission infrastructure | North America | N/A | 20-40 million | 16 days to 1-2 years | Value of Lost Load estimation | Not stated | $0.6-2.6 trillion (USD) | Not modelled | Not modelled | No |
| 2014 | Schulte in den Bäumen et al. | Electricity transmission infrastructure | Global | National assessment | Not stated | 5 months to 1 year | Multi-Regional Input-Output analysis | Not modelled | Not stated | Not stated | $3.4 trillion (USD) | Yes |
| 2014 | Schrijver et al. | Electricity transmission infrastructure | North America | N/A | N/A | N/A | Retrospective cohort exposure study with controls | Not stated | ~4% of claims are statistically associated with geomagnetic activity | Not modelled | Not modelled | Yes |
| 2017 | Forbes and St. Cyr | Electricity transmission infrastructure | England and Wales | N/A | N/A | N/A | Econometrics | N/A | N/A | N/A | N/A | Yes |
| 2017 | Oughton et al. | Electricity transmission infrastructure | USA | National assessment | 8-66% | 24 hours | Multi-Regional Input-Output analysis | Not modelled | $3-28.2 billion (USD) | $1.4-7.2 billion (USD) | $4.4-35.4 billion (USD) | Yes |

## *1989-2007*

After the 1989 geomagnetic storm a variety of researchers began to focus on the potential impacts on the power grid. This included calculating the economic costs of both asset damage, the unserved electricity load, and replacement power. Mitigation costs were first examined, with Douglas (1989) exploring investment into neutral blocking and grounding devices for the electricity transmission grid in the wake of the Quebec event, where six million people lost power (Boteler, 1991). Then later, cost information surfaced in a paper by Bolduc (2002), a researcher at the Hydro-Québec Research Institute, focusing on GIC observations in the Hydro-Québec power system. The cost of damaged equipment to Hydro-Québec from overvoltages was $6.5 million Canadian Dollars in material damages alone, among a total of $13.2 million (although this is likely to be relatively insignificant when compared to the wider economic impact). Indeed, the damages to transmission equipment were quite severe as it took many months to get some assets repaired and fully operational again.

In the same March 1989 event, utilities in the northern USA also experienced problems. Consequently, the US Department of Energy funded a piece of research at the Oak Ridge National Laboratory by Barnes and Dyke (1990) which undertook an analysis of a geomagnetic disturbance affecting the power grid in the north east of the USA for a 48-hour period. It is stated in the analysis that several transformers were damaged and removed from service during the 1989 event, in particular at the Salem nuclear power plant in the state of New Jersey where the replacement cost was reported as 'several million dollars, and the replacement energy cost was about $400,000 a day for 6 weeks, while the plant was shut down' (p3). Fortunately, replacement transformers were available, otherwise the plant could have been closed for up to a year due to the prolonged delivery times of Extra High Voltage (EHV) transformers.

In the analysis by Barnes and Dyke, a scenario is explored where 95 percent of those states served by the Northeast Power Coordinating Council (Maine, Vermont, New Hampshire, Massachusetts, New York, Connecticut, Rhode Island), as well as New Jersey and the majority of Pennsylvania, are without power due to a blackout caused by a geomagnetic disturbance. The authors use a power restoration process where 50% of the population are reconnected after 16 hours, 75% after a day and 100% after 2 days. This event was assumed to cause damage at two nuclear power plants to four single-phase power transformers. The methodology focused on estimating the Value of the Lost Load (VOLL) during the period of the blackout. VOLL is a monetary indicator expressing the costs associated with electricity supply interruption (Schröder and Kuckshinrichs, 2015), however it only captures the direct effects of electricity supply interruption, and does not factor in the multiplier impacts which can accrue throughout the economy. Following a literature review the estimated cost impacts were determined to be between $1.87 to $3.33 per unserved kWh of demand (1988 USD), for both residential, and industrial and commercial customers. Using estimates for the regional average hourly energy load, the lost load of unserved electricity was then multiplied by the lower and upper costs per kWh to obtain an estimated range.

The costs of replacement power are also calculated resulting from damaged transformers, as replacement power is assumed to be required for 12 months while new transformers are constructed, given bespoke spares are unlikely to be found. A loss of two nuclear 1100 MW power plants (operating at 65% capacity) leads to a power replacement cost of $313 million to $1.253 billion for 12 months depending on the cost per kWh. Based on these assumptions, the direct economic costs range from $3.0 to $6.1 billion. The value associated with the unserved load ranges between 79% to 89% of the cost, the electricity replacement cost was smaller at between 10-20%, and the transformer cost was negligible. Kappenman (1996) later quotes this study in an article for IEEE Power Engineering Review, where he relates these impacts to other natural hazards taking place in 1989, stating it is equivalent to Hurricane Hugo or the San Francisco earthquake.

In comparison to the March 1989 event, the October 2003 geomagnetic storm was less severe, and led to less of an economic impact. Due to ongoing maintenance works on the transmission grid in southern Sweden, the Malmö region underwent a large-scale blackout due to significant GIC. According to Pulkkinen et al. (2003), for approximately an hour about 50,000 people were left without power although this led to a relatively minor economic impact of approximately 0.5 million US dollars from the unserved electricity. The loss of power was reported to have caused significant local issues however, delaying many trains, and leaving many people stranded in elevators.

Reflecting on these studies, the economic impact assessments undertaken between 1987 and 2007 were relatively basic, focusing firstly on the cost to damaged infrastructure assets, and secondly on utilising VOLL techniques to calculate the direct economic impact. The wider economic impacts are hence not considered.

## *2008-2013*

Almost two decades after the Quebec 1989 incident a workshop was held in Washington DC, on May 22[nd] 2008 under the auspices of the US National Research Council's Space Studies Board. Bringing together industry, Federal government representatives and social scientists, the workshop focused on the potential societal and economic impacts of severe space weather, and produced an account of the event in the form of an extended summary report (see Space Studies Board, 2009). This raised the profile of the potential socio-economic impacts of space weather by examining the level of disruption to CNI. In particular, Kappenman's contribution within the report points to an estimate by the Metatech Corporation that "the total cost of a long-term, wide-area blackout caused by an extreme space weather event could be as much as $1 trillion to $2 trillion during the first year, with full recovery requiring 4 to 10 years depending on the extent of the damage" (Space Studies Board, 2009:13). This became a widely quoted figure following the event, despite little evidence being provided to support the claim. Indeed, tracing the spatial and temporal assumptions used to arrive at this figure is challenging, but due to the size of the proposed impact, one would assume this covers both direct and indirect costs to the economy. Within the report, the author makes comparative reference to the 2003 blackout that took place in the north east of the USA and Ontario, Canada (a non-space weather induced critical infrastructure failure) which affected 50 million people and led to an estimated cost of between $4 billion and $10 billion according to the US-Canada Power System Outage Task Force (2004).

A Metatech report was later prepared by Kappenman (2010) for Oak Ridge National Laboratory entitled 'Geomagnetic Storms and Their Impacts on the US Power Grid'. Essentially a vulnerability assessment, the report provided a detailed overview of the modelling and analysis undertaken. It also explored why the replacement of Extra High Voltage transformers takes such a protracted period of time. Later, a JASON report (2011) was specifically tasked with assessing the worst-case scenario put forward by Kappenman, which proposed that the US transmission grid could undergo catastrophic damage, leaving millions without essential electricity services for anywhere in between a few months to a few years. Tasked by the US Department of Homeland Security one of the primary objectives of the study was to assess 'the plausibility of Mr Kappenman's worst-case scenario' (JASON, 2011:1) which included (i) catastrophic damage to >300 Extra High Voltage transformers, (ii) 130 million people without power for several years, and (iii) a $1-2 trillion (USD) economic impact.

However, the findings of the report highlight some key issues with the original analysis. Firstly, both data and algorithms used in the analysis were stated to be proprietary and were therefore unavailable for examination by the JASON investigators, hence the report emphasised that national policy should not be based on methods which are not fully transparent and available to key decision-makers. Secondly, the nature and characteristics of the Extra High Voltage transformers being analysed are not well known, as exemplified in the analysis by the extrapolation of small samples of data. Therefore, it is incredibly challenging to robustly quantify how many may fail when exposed to different levels of GIC. Finally, experiences of other transmission infrastructure systems exposed to high GIC have not necessarily experienced catastrophic damage, for example in both Quebec and Finland, making the authors doubt the suggested impacts in the worst-case scenario. The study concludes by stating:

*"We agree that the U.S. electric grid remains vulnerable but are not convinced that Kappenman's worst-case scenario [26] is plausible, i.e. that a severe solar storm will probably destroy up to 300 EHV transformers, leaving as many as 130 million people without power for years while replacement transformers are manufactured and installed."* (JASON, 2011:64-65)

This is an important turning point in the narrative associated with the economic impacts of space weather, as it raised doubts over there being extremely long restoration periods (and hence trillion dollar impacts). However, the idea that a space weather event could cause *trillions* of lost dollars was alarming for many of risk bearers who consequently undertook their own scenario-based analysis. One example relates to the insurance industry who not only insure the physical assets of many critical infrastructure operators, but also cover property, casualty and supply chain insurance more generally throughout society and the economy. Hence, the worry is that if this type of tail event actually took place it could be enough to put many insurers out of business, especially as existing vehicles for spreading risk may not address space weather.

Consequently, Lloyd's of London (2013) commissioned Atmospheric and Environmental Research to undertake an assessment of the risk to the North American grid. The analysis can be seen as a primer for corporate risk managers to begin to understand how CMEs can drive geomagnetic storms on Earth, initially covering the risk factors that may increase exposure, such as geomagnetic latitude, ground conductivity, coastal effects and transmission system characteristics. The assessment uses Carrington-level geomagnetic storm simulations for the geomagnetic field and then relates these to local ground conductivity structure (see Wei et al. 2013). A power grid model is then used to assess the level of GIC flowing through transformers (identified using commercially available transformer data) to assess transformer vulnerability due to thermal heating. Grid instability impacts are therefore not modelled. Using estimated transformer age distributions and temperature information, outage scenarios are estimated at the county level.

The appendix details how the economic costs are derived, using a similar VOLL method to Barnes and Dyke (1990), by calculating the cost of the unserved electricity in 2001 USD. A linear restoration is assumed using $2.00/kW, $19.38/kW and $8.40/kW for residential, commercial and industrial users. The study findings are that for a Carrington-level storm, whereby 20-40 million people are affected

for between 16 days up to 1-2 years, the total economic cost is estimated at $0.6-2.6 trillion USD. The logic for this analysis in terms of impact zone size and restoration time is similar to the proposed Kappenman worst-case scenario, but again transformer age distributions and characteristics are essentially assumed. In the recommendations of the JASON (2011) report, it is emphasised that the actual transformer asset data should be collected to enable comprehensive simulation of the entire grid. Although improvements were made in the analysis of electricity transmission infrastructure in the period between 2008-2013, key disagreements arose due to diverging views on how our engineered systems would respond. In some cases this led to some very large economic impacts, but with little detail on the economic methodology used to justify the proposed numbers.

## *2014-2017*

In 2014, a multi-disciplinary group of scientists and economists published a paper on how severe space weather can disrupt global supply chains (Schulte in den Bäumen et al. 2014). The key contribution of the analysis is that for the first time a more robust economic methodology was applied to space weather impact assessment, superior to the previous method of roughly calculating the lost value of the unsupplied electricity load.

None of the previously published estimates considered global impacts to international trade, therefore this was an important step forward by utilising standard macroeconomic methods. The methodology applied, known as input-output economics, is a field of analysis developed by Wassily Leontief in the late 1930s and awarded the Nobel Prize in Economic Science in 1973 (see Miller and Blair, 2009 for a comprehensive overview of the method). Importantly, the techniques provide a formal framework for analysing interindustry transactions (often monetary flows), and allows one to model the impacts of changes within the economy, both directly and indirectly. This method has been a frequent workhorse used to understand the potential economic ramifications of infrastructure failure (see Haimes and Jiang, 2001; Anderson et al. 2007; Leung, 2007; Setola et al. 2009; Pant et al. 2011; Pant et al. 2014; Jonkeren and Giannopoulos, 2014). In this example, Schulte in den Bäumen et al. (2014) firstly apply a physical model calibrated to the latitudinal (80°) and longitudinal (8°) width of the auroral electrojet, and applied a set of scenarios across different continents, in both the northern and southern hemisphere, for a storm similar to the Quebec 1989 event. This was the first time a physical model had been coupled with a global macroeconomic economic model, making a key contribution to the literature. Moreover, a set of Multi-Regional Input-Output (MRIO) tables were utilised called Eora, developed by the co-author Lenzen, which depending on the country covers between 25-400 industrial sectors for a total of 187 countries of the world. This represents 99.99% of global trade. A total grid shutdown is then modelled in those countries within the storm impact zone footprint.

Importantly, the authors make the key point that as all economic assessments to date have not considered indirect trade effects, they have not included domestic and international supply chain linkages in the estimates. This leads to a total economic loss in the USA of $25 billion USD per day, similar to the Lloyd's of London estimate of approximately $30 billion USD per day. Moreover, a Carrington-level event taking place over North America was estimated to reach $1.2 trillion USD over five months. Indeed, in the discussion the authors' state that a 'severe space weather event could be the worst natural disaster in modern history with global cost estimates to be over 5% of world Gross Domestic Product (GDP) and impacts reaching across every industry and every segment of society" (p2756). Indeed, the results of the paper indicate that the total economic impact could be $3.4 trillion USD over a year which is approximately 5.6% of global GDP.

In another key contribution, Schrijver et al. (2014) analyse a novel dataset of insurance claims belonging to the insurance company Zurich. For January 2000 up to December 2010, 11,242 insurance claims were analysed for equipment loss and affiliated business interruption for corporates located in North America. The key finding was that on days where there was elevated geomagnetic activity, the claim rate increased by approximately 20%, for the top 5% of most active days. Overall, claim rates were elevated by approximately 10% for the top third of the most active days, when ranked by the maximum variation in Earth's geomagnetic field. This finding suggests that large-scale geomagnetic activity causing variations in the quality of power provided, can induce equipment faults in electrical and electronic devices, potentially leading to an estimated 500 additional claims on average per year. Importantly, this analysis uses data from a relatively low-activity period for space weather, so this number could rise if we experienced a particular increase in solar activity. The financial implications of these claims are substantial. Indeed, they are unlikely to be related to space weather, by both the firms affected and the wider insurance industry, demonstrating that many of the cost impacts go unattributed. Research by Forbes and St. Cyr (2008; 2012; 2017) also focuses on day-to-day space weather impacts on power grids, which can still pose an challenge for network operators, potentially incurring substantial operational costs.

One particular problem with the level of analysis to date is the challenge of reconciling global or continental-scale geophysical activity with the local impacts on critical infrastructure. Indeed, while the use of macroeconomic modelling methodologies is an important step forward for estimating total economic impact (e.g. Schulte in den Bäumen et al. 2014), the local impacts of critical infrastructure failure on firms, labour and value-added activity is lost. Within the USA, there is such heterogeneity among different states, particularly in industrial composition, modelling impacts at the macroeconomic level can produce quite coarse results. In a study by Oughton et al. (2017) an assessment of electricity transmission failure due to space weather is undertaken focusing on the USA. Different scenario-based storm footprints are tested to explore the variation in direct and indirect economic impacts, as illustrated in Figure 4. The weighted population centroid of each state is represented by a black dot in this graphic, with the state being included in each scenario if each centroid falls within a set of electrojet footprints. The full paper provides further detail. These scenarios are exploratory in nature, with S1 being the most probable, and S4 being highly improbable.

Figure 4 Tested blackout zones, customer disruptions and daily direct economic impact. S1-S4 represented four scenarios reported in Oughton et al. 2017.

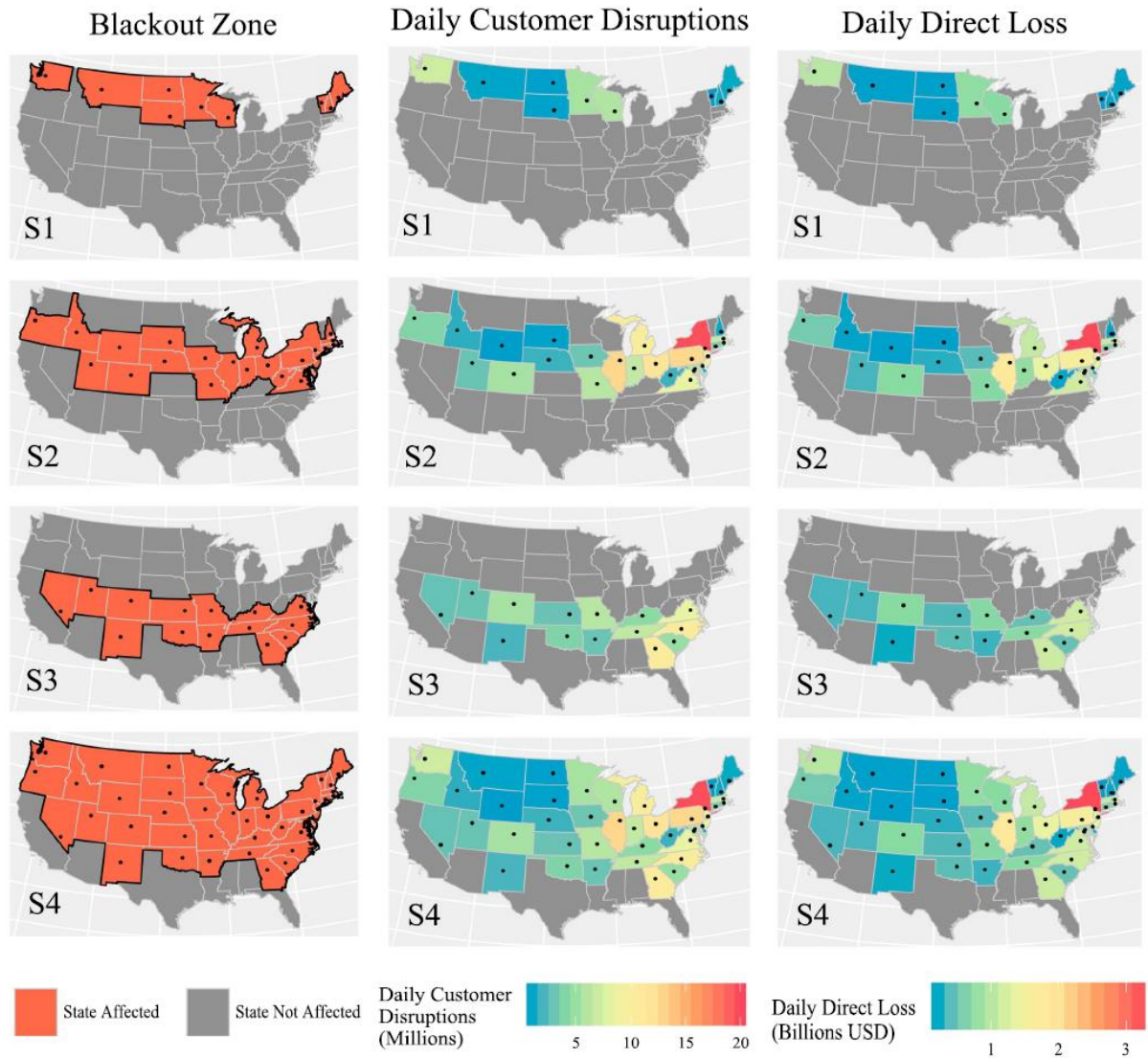

The methodology is similar to the Schulte in den Bäumen et al. (2014) paper in that it utilises MRIO data to quantify domestic and international supply chain impacts, although it does build on a different dataset called the World Input Output Database (WIOD). However, the method is supplemented by using state-level GDP output data for 20 industrial sectors and uses this to drive national shocks in the macroeconomic model, allowing the local heterogeneity in economic activity to be quantified. Due to disagreement within the scientific and engineering communities on the length of potential power outages, the analysis focused on the economic impact for a 24-hour period. Depending on the storm footprint tested, the total economic impact ranged from $6.2-41.5 billion USD per day to the US economy.

Reflecting on these analyses, both Oughton et al. (2017) and Schulte in den Bäumen et al. (2014) approximately scale the impact footprint based on the latitudinal and longitudinal width of the electrojet. However, recent research by Ngwira et al. (2015) and Pulkkinen et al. (2015) suggests that zones of extreme activity have significantly smaller footprints, with substorms taking place in localised areas within the electrojet, indicating these analyses could overestimate impacts. Like much of the literature to date, the shortcomings of these analyses are that they lack robust scientific and engineering inputs, hence paying little attention to ground conductivity and underlying grid structure. Indeed, it exemplifies the fact that many analyses on the economic impacts of space weather have lacked the level of rigour desired by the scientific and engineering community. Now that the literature over the past three decades has been critiqued, the strengths of different approaches will be reflected on, by providing an ontology for assessing the economic impacts of space weather in future research.

# An ontology for assessing the economic impacts of space weather

Reflecting on the relative strengths and weaknesses of different approaches, this section proposes how future research could progress. Assessing the economic impacts of natural hazards due to critical infrastructure failure is a well-developed field, therefore there is a degree to which one does not need to 'reinvent the wheel' in this case (see Ouyang, 2014 for a comprehensive overview on modelling infrastructure systems, including economic methods). However, the economic analysis of space weather is yet to bring together the strengths of different approaches to comprehensively assess the total impacts; information which is vitally needed by decision-makers in government and industry to help inform cost-benefit assessments for resilience.

Firstly, the actual economic costs of space weather have been calculated in a variety of ways in the literature. Many studies have focused purely on a single type of cost, such as damage to network assets, the direct cost of unserved electricity or the wider economic impacts via supply chains. No study has yet convincingly quantified the implications of each of these economic costs, despite this being essential to gain a comprehensive understanding. Moreover, some have called for further work to place a greater focus on assessing both direct and indirect losses (see Eastwood et al. 2017:213), but this activity should not be undertaken in isolation. Indeed, the conjecture proposed here is that mitigation costs should also be considered in order to enable the type of resilience analytics actually needed to support real decisions (see Rose, 2017 for further background on the benefit-cost analysis of economic resilience actions).

Finally, the economic impacts have sometimes been ambiguously defined in the literature, and therefore it is important to identify to whom they relate. For example, there could be costs to infrastructure network operators as the service provider, firms as the key actors carrying out production activities in the wider economy, and also households as consumers of final goods and services. Based on the ideas articulated here, Table 3 illustrates the different types of economic costs associated with space weather, and which stakeholders bear them.

Table 3 The economic costs associated with space weather

| Entity | Cost type | Type of damage or mitigation costs | | |
|---|---|---|---|---|
| Infrastructure network operator | Direct | Damage to assets | Idle resources | Lost sales of electricity |
| | Indirect | Delayed supply of replacement assets | Increased cost of replacement assets | |
| | Mitigation | Operational mitigation measures | Asset upgrades and blocking devices | System upgrades including islanding |
| Commercial and industrial customers | Direct | Production downtime | Delayed scheduling of activities | |
| | Indirect | Delayed supply of upstream goods and services | Delayed delivery of downstream goods and services | |
| | Mitigation | Backup power | Inventory stockpiling | Higher electricity tariffs to fund network upgrades |
| Households | Direct | Lost leisure time | Property and casualty damages | |
| | Indirect | Constrained consumer spending due to unavailable goods and services | Price increases due to short supply of goods and services | |
| | Mitigation | Backup power | Higher electricity tariffs to fund network upgrades | |

As many studies have focused on different types of costs, this has often led to large divergence in proposed estimates, for example from the equipment damage costs reported after the Quebec event, to scenario-based estimation of the wider economic impacts. However, this is exacerbated by the high level of uncertainty associated with the spatial and temporal impacts of potential space weather events.

Within the catastrophe modelling paradigm, techniques for addressing this already exist with regard to quantifying the potential impact of earthquakes, hurricanes and other natural hazards by capturing both primary and secondary causes of uncertainty. Primary uncertainty relates to the size and location of the storm impact zone for a space weather event and the problems in predicting this. For example, the initial impact with the most intense substorm could affect North America, Europe or East Asia depending on the daily rotation of the Earth. But it also reflects that fact that we have quite poor time-series data and only a partial understanding of the physical processes driving space weather. Subsequently, the variability in local intensity, asset damage and different economic impacts pertain to the level of secondary uncertainty. Figure 5 illustrates an ontology for assessing the economic impacts of space weather, recognising both primary and secondary drivers of uncertainty, and potential direct and indirect economic impacts.

Figure 5 Ontology for assessing the economic impacts of space weather

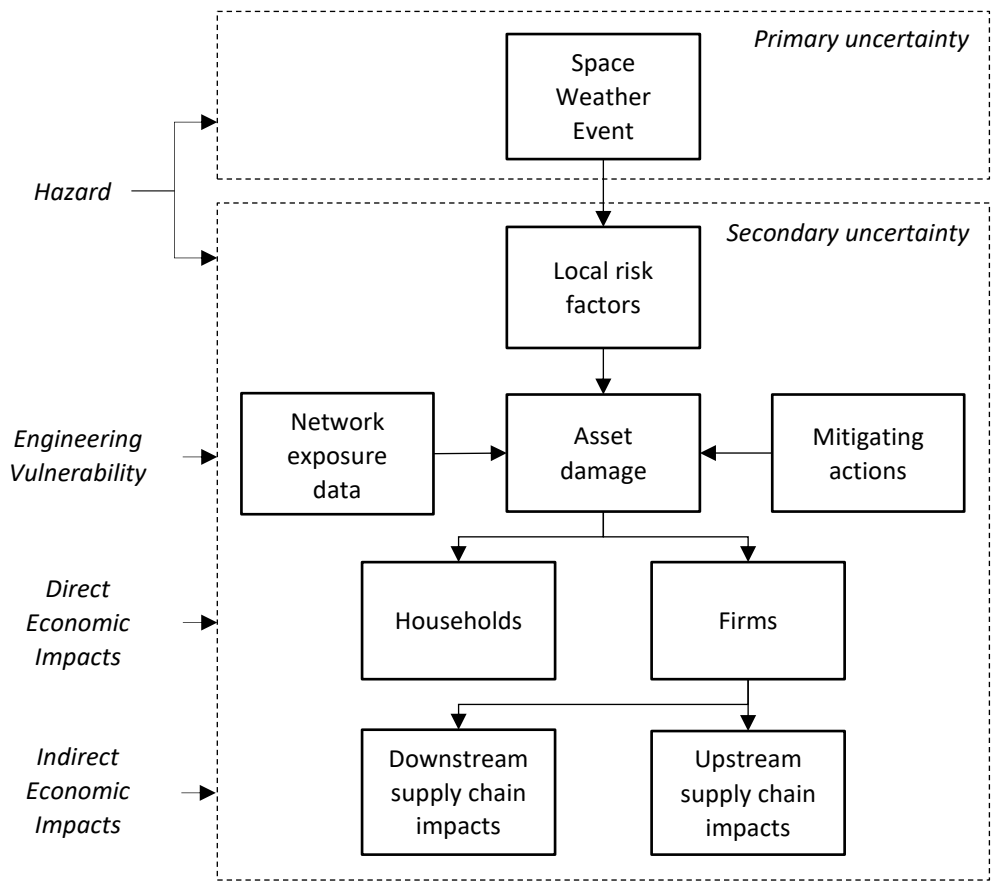

Local risk factors can either increase or decrease the potential risk. As already identified, ground conductivity, geomagnetic latitude and coastal proximity are important geophysical parameters that affect the amount of GIC produced, although it is very hard to predict this in reality. Many economic assessments have not utilised actual infrastructure network data leading to an overly simplistic estimation of the impacts, hence Figure 5 highlights the need to use network exposure data. This has the advantage of being able to anchor vulnerability and risks to assets within a spatial context which is vital for robustly quantifying potential economic costs to the network operator, as well as households and firms. Damage to assets can be quantified as replacement or repair costs to estimate the economic impact on the network operator.

A variety of mitigation options are available which have the potential to limit damage to infrastructure assets. In generally, mitigation options are quite different from other natural hazards due to the characteristics of space weather. Earthquake risk requires assets to be made more structurally robust from increased physical vibration, and similarly flooding risk requires assets to be elevated or surrounded by embankments. Yet with space weather, it is specific technological components within infrastructure assets that can be most affected and require hardening or protecting in other ways, such as the winding conductors of EHV transformers. As well as physical installations of mitigation equipment (e.g. in the power grid), and enhancing space weather forecasting and early warning systems, there are also a variety of operational mitigation measures that can be implemented across the network. These include (i) increasing spinning reserve, (ii) cancelling maintenance to make as many lines as possible operational, and (iii) introducing islanding and microgrid configurations. However, many operational measures are highly dependent on satellite capabilities providing early warning that a space weather event is actually taking place. Therefore, ensuring this capability remains available is crucial for ensuring operational awareness and protecting critical infrastructure.

Finally, the wider economic impacts can be assessed as taking place to both households and firms. Where asset data for infrastructure networks can be quite challenging to obtain, population and business statistics are relatively ubiquitous across national statistical bureaus. Often these statistics are available at a high level of granularity, and if no specific network structure is available then can be spatially joined to the closest network asset using open-source software. Importantly, firms are often part of both domestic and international supply chains and therefore these impacts need to be quantified for a comprehensive cost assessment. Indeed, upstream economic impacts take place indirectly because firms directly affected by critical infrastructure failure buy fewer goods and services. Moreover, downstream economic impacts also accrue as firms directly affected by the event are unable to sell their goods and services to other firms which may often be critical for their production processes. In the case of final products, consumers may not be able to purchase their desired goods and services leading to reduced aggregate demand in the economy.

Each of the areas within Figure 5 have been more or less addressed within the literature already, however they have often been done so in isolation, preventing the true economic cost of space weather being explored. The concepts of primary and secondary uncertainty, while a cornerstone of $21^{st}$ century catastrophe modelling, have yet to percolate into the economics of space weather. This is partly because the level of analysis to date has been relatively piecemeal, with few focusing their attention on this topic for a prolonged period which inevitably means there is little consistent analysis, limited incremental refinement of assessment models and no examination of the sensitivities of assumption sets. These are all areas relevant for future research. Conclusions will now be made.

## Conclusions

This article has tracked how the economic impact assessment of space weather has evolved over the past three decades, focusing mainly on the risks posed to electricity transmission infrastructure. Although a number of key contributions have taken place, the economic analysis of this natural hazard still lags behind other natural catastrophe threats such as hurricanes or earthquakes, partially due to the low probability, high impact nature of extreme space weather. While we have not experienced a Carrington-sized event in recent times, we do regularly experience moderate space weather activity. Yet, often we may fail to properly attribute problems induced by this hazard, partially due to a lack of awareness, but also because failure of infrastructure assets and other electrical equipment may take place in the days, weeks or months following an event. Despite space weather continuing to receive growing attention across industry and government, much progress needs to be made in trying to understand the degree of secondary uncertainty associated with the potential economic impacts of space weather. This includes exposure, vulnerability and potential damage to key infrastructure assets.

As well as charting the evolution of the literature, an ontology was also proposed for the comprehensive economic assessment of space weather. This has been based on the relative strengths of certain methods and motivated by the need for this evidence in decision-making regarding mitigation measures. Hitherto, key studies have independently addressed the economic impact to key entities involved, whether that be the infrastructure network operator, the wider supply chain impacts on firms or players in the insurance market. However, future work needs to bring together the potential detrimental cost implications for each of the entities within the economy, and assess these impacts in relation to investment in mitigation measures, allowing a more advanced understanding of the cost-benefit trade-offs associated with this natural hazard.

A critical issue with many of the assessments carried out over the past three decades is that many completely overlook the actual physical impacts on infrastructure assets and the topology of the infrastructure networks, often relying far too heavily on qualitative assumptions about CNI vulnerability. This is problematic because it gives rise to a relatively weak evidence base. Although data on infrastructure assets and the network topology can sometimes be challenging to obtain, the tools are available to explore the sensitivity of the uncertainty associated with CNI vulnerability. Indeed, the JASON (2011) report makes reference to this, in that researchers should be making more use of simulation techniques to increase the evidence base on this matter. Part of this also relates to data sharing which can be a perennial issue, especially when dealing with asset information which is potentially business sensitive.

In terms of best practice, as efforts on this matter have historically been relatively piecemeal with few authors publishing more than once on the economic impacts of space weather, there has been little explicit exploration of model sensitivities, including in relation to different assumption sets. Increasingly open-source model code is becoming the gold standard, although this review found little evidence of researchers moving in this direction. Given the importance of this under-researched hazard, and the need for national policy to be based on fully transparent data, models and tools, making these steps would be a worthy contribution to the economic impact assessment of space weather.

## Acknowledgements

Edward Oughton would like to express his gratitude to the UK Engineering and Physical Science Research Council as this research was financially supported under grant EP/N017064/1: Multi-scale InfraSTRucture systems AnaLytics.